\documentclass[twocolumn,showpacs,superscriptaddress, preprintnumbers , amsmath,amssymb,letterpaper, prb]{revtex4-1}

\usepackage{scrextend}
\usepackage{graphicx}
\usepackage{dcolumn}
\usepackage{bm}
\usepackage{amsmath}
\usepackage{multirow}
\usepackage{textcomp}

\newcommand{\mean}[1]{\mbox{$\langle{#1}\rangle$}}

\usepackage[colorlinks=true]{hyperref}
\usepackage{amsmath}
\usepackage{amssymb}
\usepackage{array,booktabs}
\usepackage{graphicx}
\usepackage{amsbsy}

\begin{document}
\title{Tailoring of an Electron-Bunch Current Distribution via Space-to-Time Mapping \\
of a Transversely-Shaped Photoemission-Laser Pulse}
\author{A. Halavanau} 
\altaffiliation[Present address: ]{SLAC National Accelerator Laboratory, Menlo Park, CA 94025, USA}
\affiliation{Northern Illinois Center for Accelerator \& Detector Development and Department of Physics, Northern Illinois University, DeKalb, IL 60115, USA} 
\affiliation{Fermi National Accelerator Laboratory, Batavia, IL 60510, USA}
\author{Q. Gao} 
\affiliation{Argonne Wakefield Accelerator, Argonne National Laboratory, Lemont, IL 60439, USA} 
\affiliation{Accelerator laboratory, Department of Engineering Physics, Tsinghua University, Beijing, China} 
\author{M. Conde} 
\affiliation{Argonne Wakefield Accelerator, Argonne National Laboratory, Lemont, IL 60439, USA} 
\author{G. Ha} 
\affiliation{Argonne Wakefield Accelerator, Argonne National Laboratory, Lemont, IL 60439, USA} 
\author{P.  Piot} 
\affiliation{Northern Illinois Center for Accelerator \& Detector Development and Department of Physics, Northern Illinois University, DeKalb, IL 60115, USA} 
\affiliation{Fermi National Accelerator Laboratory, Batavia, IL 60510, USA}
\author{J. G. Power} 
\affiliation{Argonne Wakefield Accelerator, Argonne National Laboratory, Lemont, IL 60439, USA} 
\author{E. Wisniewski} 
\affiliation{Argonne Wakefield Accelerator, Argonne National Laboratory, Lemont, IL 60439, USA}

\begin{abstract} 
Temporally-shaped electron bunches at ultrafast time scales are foreseen to support an array of applications including the development of small-footprint accelerator-based coherent light sources or as probes for, e.g.,  ultrafast electron-diffraction. We demonstrate a method where a transversely-segmented electron bunch produced via photoemission from a transversely-patterned laser distribution is transformed into an electron bunch with modulated temporal distribution. In essence, the presented transformation enables the mapping of the transverse laser distribution on a photocathode surface to the temporal coordinate and provides a proof-of-principle experiment of the method proposed in  Reference~ [\href{https://journals.aps.org/prl/abstract/10.1103/PhysRevLett.108.263904}{W. S. Graves, et al.  Phys. Rev. Lett. {\bf 108}, 263904 (2012)}] as a path toward the realization of compact coherent X-ray sources, albeit at a larger timescale. The presented experiment is validated against numerical simulations and the versatility of the concept, e.g. to tune the current-distribution parameters, is showcased. Although our work focuses on the generation of electron bunches arranged as a temporal comb it is applicable to other temporal shapes. 
\end{abstract}
\date{\today}

\maketitle

\section{introduction}
Spatio-temporal control of bright electron beams has a wide array of applications including compact-footprint accelerator-based light sources~\cite{graves}, 
ultra-fast electron probe setups~\cite{zewail}, and the possible development of efficient high-energy accelerator based on beam-driven wakefield-acceleration 
techniques~\cite{zholents}. One recurrent class of distribution that could benefit most of these applications is the case of a bunch train where the beam is organized as ``microbunches" in time with a characteristic duration below $\sim 1$~ps. Owing to the fast time scale involved, this type of distribution is challenging to compactly achieve in conventional accelerator beamlines where the three degrees of freedom are decoupled. The last decade has witnessed the development of accelerator beamlines where the phase-space coordinates associated with different degrees of freedom can be exchanged~\cite{Brinkmann,FBpiot,EmmaEEX1,PiotPRABshaping}.  

The present work combines one type of such a beamline with a photoemission source to control the {\em temporal} distribution of the photoemitted electron beam using an optical {\em transverse}-shaping technique to adequately alter the emission-triggering laser distribution on the photocathode surface. The considered space-to-time mapping  beamline is a transverse-to-longitudinal phase space exchanger  (TLEX)~\cite{EmmaEEX2}. The principle of the TLEX can be best understood by considering the phase space coordinate $(x,p_x,y, p_y,\zeta, p_z)$ where $(x,y,\zeta)$ is the electron position with respect to a reference point (henceforth taken to be the electron bunch barycenter) and $(p_x,p_y,p_z)$ is the corresponding momentum. In our convention, the beam propagates along the $z$ direction so that $p_z\gg p_x, p_y$. In addition, we define the relative longitudinal coordinate $\zeta\equiv z-\mean{z}$ where $\mean{.}$ stands for the statistical averaging over the bunch distribution. Here we note, under the above assumptions, that $\zeta \simeq c t$ for a relativistic beam where $c$ is the velocity of light and $t$ the time coordinate defined with respect to the bunch barycenter. Therefore the longitudinal coordinate $\zeta$ is often referred to as the temporal direction. 

A common TLEX setup consists of a  transverse-deflecting cavity (TDC) operating on the TM$_{110}$ at zero crossing mode flanked by two identical ``dogleg" beamline each composed of two dipole magnets with the same deflecting angle but with reversed polarities. Under a simple linear model of the TLEX beamline and treating the horizontally-deflecting TDC as a thin lens, the coordinate of an electron in the four-dimensional phase space (using angle instead of momentum) downstream of the TLEX is $\pmb u= R \pmb u_0$, where $\tilde{\pmb u} \equiv (x,x'\equiv p_x/p_z, \zeta, \delta \equiv p_z/\mean{p_z}-1)$, with $\tilde{.}$ referring to the transpose operator,  $\pmb u_0$ is the corresponding vector upstream of the TLEX, and the $4\times4$ $R$ transfer matrix~\cite{EmmaEEX1} is given by
\begin{eqnarray}
R=
\begin{pmatrix} 
0  & A \\
-A^{-1} & 0 
\end{pmatrix} 
\mbox{~with~} 
A\equiv
\begin{pmatrix} 
-\frac{L}{\eta} & \eta-\frac{\xi L}{\eta} \\
\frac{-1}{\eta} & -\frac{\xi}{\eta}  
\end{pmatrix}. 
\end{eqnarray}
The parameters $\eta \equiv \frac{\partial x}{\partial \delta_0}\big|_{x=0}$ and $\xi \equiv \frac{\partial \zeta}{\partial \delta_0}\big|_{\zeta=0}$ are respectively the horizontal and longitudinal dispersion functions generated by one dogleg and $L$ the resulting path length. The block-antidiagonal nature of the transfer matrix is achieved when the cavity deflecting strength $\kappa\equiv \frac{2\pi}{\lambda_{rf}} \frac{e V_{\perp}}{cp_z}$ satisfies $\kappa=-1/\eta$ where $\lambda_{rf}\simeq 23$~cm  is the RF wavelength, $e$ is the electronic charge, and $V_{\perp}$ the deflecting voltage.
 It should be noted that the TLEX phase-space exchange occurs within the $(x,x', \zeta, \delta)$ phase space so that the $(y,y')$ vertical phase space remains uncoupled and is therefore ignored in our treatment. Owing to the block-anti-diagonal nature of the transfer matrix $R$, the initial horizontal phase-space $(x,x')$ coordinates are mapped to the longitudinal phase space $(\zeta, \delta)$  and vice versa. It was suggested~\cite{PiotPRABshaping}  and experimentally shown that such a feature could be taken advantage of to tailor the longitudinal (current) distribution of an electron beam~\cite{SunPRL,Maxwell,HaPRL}. 

In most of the previous work, the current distribution is experimentally controlled by locating an interceptive mask upstream of the TLEX. In the present work, a simple transverse laser-shaping technique controls the transverse phase-space distribution upstream of the TLEX which is then mapped to the longitudinal phase-space distribution downstream of the TLEX. We especially demonstrate the transfer of the projected transverse laser distribution along one of the transverse directions at the photocathode surface to the current distribution of the electron bunch downstream of the TLEX. 
 \begin{figure}[hhhh!!!]
 \includegraphics[width=0.99\linewidth]{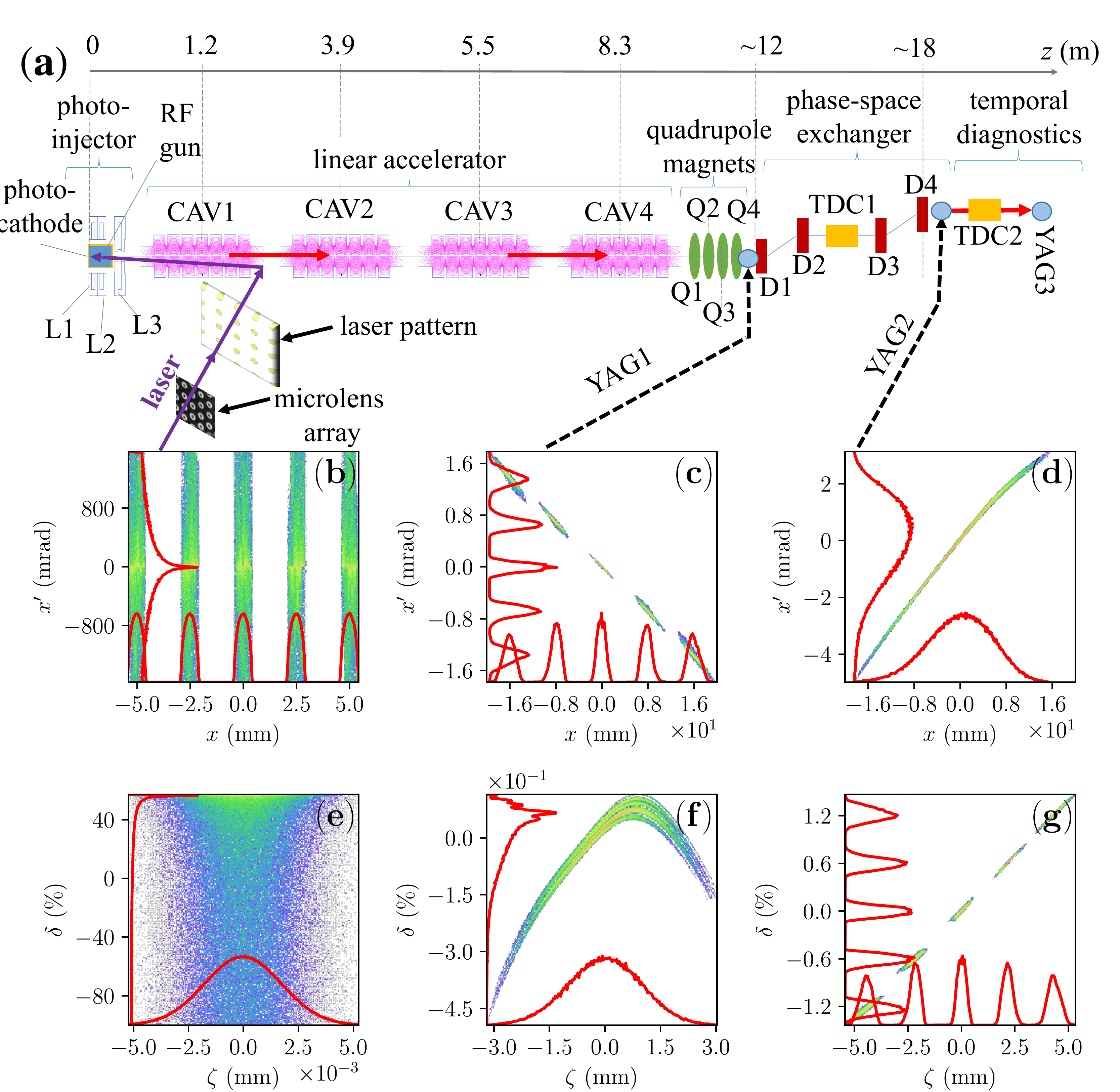}
\caption{\label{fig:AWA}Overview of the AWA facility (a). The RF-gun includes three solenoids (L1-3). The ``CAV" labels correspond to the accelerating cavities,  ``Q" and ``D" respectively refer to the quadrupole and dipole magnets, and ``TDC" to the transverse deflecting cavities. The TLEX beamline comprises the elements D1, D2, TDC1, D3, and D4. The density plots (b-d) and (e-g) respectively represent the simulated transverse and longitudinal phase spaces at the cathode surface (b,e), and YAG1 (c,f) and YAG2 (d,g) locations. The initial laser pattern is idealized as a $5\times 5$ array of uniform beamlets. }
\end{figure}

We specifically explore the formation of a modulated current profile obtained from a shaped laser distribution consisting of an array of transversely-separated beamlets generated via shaping of the laser using a micro-lens array (MLA) system; see Ref~\cite{mlaabcd,halavanauMLA}. It should, however, be stressed that the developed technique is general and could in principle be adapted to tailor arbitrary current distributions using more elaborate laser-shaping methods using, e.g.,  a digital micro-mirror device~\cite{sliPRAB}.

\section{Experimental setup}
The experiment was performed at the Argonne Wakefield Accelerator (AWA) facility~\cite{manoel}  diagrammed in Fig.~\ref{fig:AWA}(a). In brief, the AWA accelerator consists of a radio-frequency (RF) photoemission source consisting of a 1+1/2-cell 1.3-GHz standing-wave (SW) resonant cavity (dubbed ``RF gun") operating on the TM$_{010,\pi}$ mode. The RF gun is surrounded by three solenoidal lenses (L1, L2, and L3) to control the beam size. The electron bunches are formed by impinging a $3.4$-ps [root mean square (RMS)] long ultraviolet (UV, $\lambda=248$~nm) pulse on a Cesium Telluride (Cs$_2$Te) photocathode. The electron bunches reach an energy of $\sim  8$~MeV downstream of the RF gun and are then injected in a linear accelerator (linac) consisting of four 7-cell SW resonant RF cavities operating on the TM$_{010,\pi}$ mode at 1.3~GHz. The final beam energy was $E=46.7 \pm 1$~MeV during our experiment. Downstream of the linac section, four quadrupole magnets provide control over the transverse beam parameters upstream of the TLEX beamline.

 The TLEX beamline consists of two $L\simeq 2.7$~m-long doglegs each introducing a longitudinal and horizontal dispersion of $\xi=-0.30$~m and $\eta=-0.89$~m resulting in a transverse-to-longitudinal demagnification of $|R_{31}|\equiv \big|\frac{\partial \zeta}{\partial x_0}\big| = \big|\frac{\xi}{\eta}\big|\simeq 0.28$. The TLEX beamline is followed by a vertically deflecting cavity (TDC2) which, combined with a downstream cesium-doped Yttrium aluminum garnet (YAG)  scintillating screen (YAG3) provides a direct measurement of the final current distribution~\cite{gaoTDC}. The accelerator beamline also includes YAG screens upstream (YAG1) and downstream (YAG2) of the TLEX; see Fig~\ref{fig:AWA}(a).

\section{Generation of temporally modulated beam}
The MLA system used to tailor of the laser distribution consists of a pair of $10\times 10$--mm$^2$ square array consisting of spherical microlenses with effective focal lengths of $5.1$~mm and pitch of $300$~\textmu m along both directions. The MLAs were installed on a rotatable mount to remotely control the array orientation such to ($i$) pre-compensate for the rotation arising from the Larmor precession as the beam propagates through the RF-gun solenoidal lenses and ($ii$) to dynamically control the electron beam's longitudinal modulation downstream of the TLEX by varying the rotation angle of the pattern upstream of the TLEX. 

A typical transverse pattern generated by the MLA appears in Fig.~\ref{fig:AWA}(a) and is used as input for the numerical simulations of the process. All the supporting numerical simulations presented in this paper are performed with the beam-dynamics program {\sc impact-t}~\cite{impactt} which represents the electron bunch as a collection of 200,000 macroparticles (each representing $\sim 10,000$ electrons) and tracks the corresponding distribution in an user-specified accelerator beamline described by tabulated electromagnetic field maps. The {\sc impact-t} software also models space-charge effects using a particle-in-cell algorithm to solve Poisson's equation in the beam's rest frame before computing the resulting electromagnetic fields in the laboratory frame and applying the corresponding Lorentz force on each macroparticle. Figures~\ref{fig:AWA}(b-d) and (e-g) respectively displays snapshots of the simulated horizontal and longitudinal phase-space evolution along the AWA beamlines for an optimized set of beam parameters. The simulation demonstrates the process wherein the modulated transverse phase-space [Fig.~\ref{fig:AWA}(c)] is exchanged with the smooth longitudinal phase space [Fig.~\ref{fig:AWA}(f)]  to yield a temporally-modulated current distribution [Fig.~\ref{fig:AWA}(g)]. .
\begin{figure}[hhh!!!]
 \includegraphics[width=0.99\linewidth]{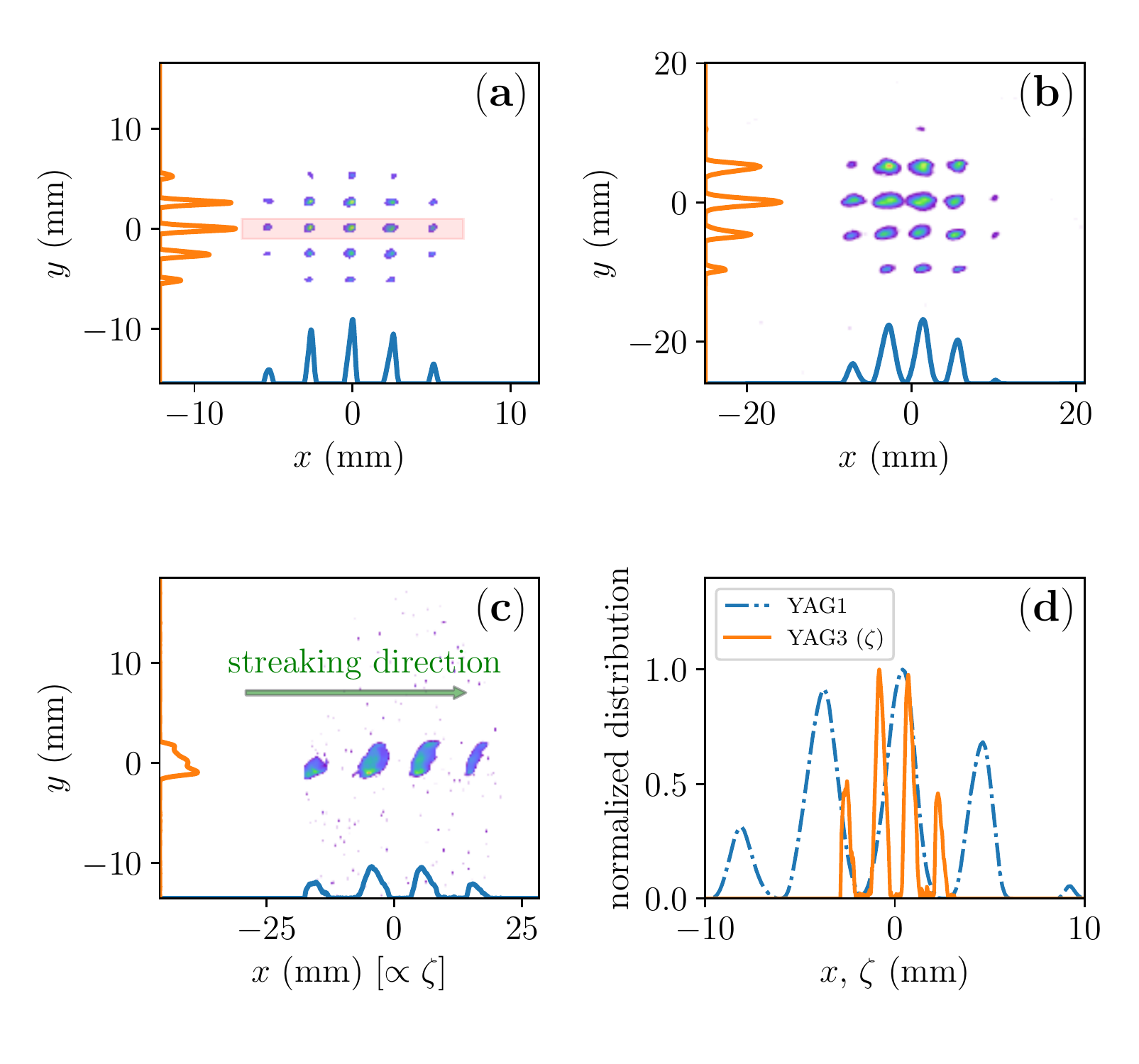}
 \caption{\label{fig:trandist} Initial (unrotated) ultraviolet laser distribution measured on the virtual photocathode (a) and corresponding electron-beam distributions measured at YAG1 (b) and YAG3 with TDC2 powered (c) with associated horizontal (YAG1) and longitudinal (YAG3) distributions (d). The shaded area in (a) indicates the row of beamlets selected for the quadrupole-magnet scan experiment (see text for details).}
\end{figure}

The MLA optical system was configured to produce a laser distribution consisting of a transverse array of beamlets with a pitch of $d=2.5\pm 0.2$~mm. The initial UV laser distribution recorded on the ``virtual" photocathode $-$ a one-to-one optical image of the photocathode surface $-$ appears in Fig.~\ref{fig:trandist}(a). The laser intensity was tuned to maintain the total bunch charge to $Q=250\pm 50$~pC, yielding $\sim$10 pC/beamlet, throughout the experiment. The transverse electron-bunch distribution recorded at YAG1, upstream of the TLEX beamline, confirms that with proper solenoidal-lens settings the multibeam pattern can be imaged in the $(x,y)$ plane~\cite{halavanauMLA}; see Fig.~\ref{fig:trandist}(b). It should be noted that the imaged pattern is generally rotated as the beam reaches the location of YAG1 ($z_{\mbox{\tiny YAG1}}$) by the Larmor angle $\vartheta_L = \int_0^{z_{\mbox{\tiny YAG1}}} \frac{eB(z) }{2 c \gamma(z)\beta(z)}dz $ acquired as the bunch experiences the axial magnetic field  produced by the solenoidal lenses surrounding the RF gun (here $\gamma$ the Lorentz factor and $\beta\equiv \sqrt{1-1/\gamma^2}$). 
For the solenoidal lens settings used in the experiment $\vartheta_L \simeq 60^{\circ}$. Consequently,   the pattern laser distribution is rotated by an angle $-\vartheta_L$ [compared to what displayed in Fig.~\ref{fig:trandist}(a)  for clarity] in order to obtained an non-rotated distribution at YAG1 [Fig.~\ref{fig:trandist}(b)]. 

In order to properly match the multi-beam array parameters so to provide a longitudinally-modulated electron beam downstream of the TLEX, four quadrupole magnets (Q1-4; see Fig.~\ref{fig:AWA}) were available upstream of the TLEX~\cite{haPRAB}. Nominally, 
an imaging was obtained when powering only Q1 and Q4 with respective magnetic-field gradient of $B'_1\simeq 0.7$~T/m and $B'_4\simeq -0.7$~T/m [where the gradient is defined as $B'\equiv\frac{\partial B_x}{\partial y}=\frac{\partial B_y}{\partial x}$ with $B_{i}$ ($i=x,y)$ being the transverse magnetic field components].   Under these 
settings, the temporal bunch distribution downstream of the TLEX consists of five microbunches displayed in Fig.~\ref{fig:trandist}(c) and in qualitative agreement with the simulated final longitudinal-phase-space distribution; see Fig.~\ref{fig:trandist}(d) thereby confirming that the beamline realizes an imaging of one of projections associated with the laser transverse distribution on the longitudinal (temporal) coordinate. 

It should be noted that due to the beamline limited aperture,
non-optimized transport, and possible beam misalignment
the side beamlets are not fully transmitted resulting in a
decrease of their charges and eventually one of the
beamlets not being transmitted downstream of the TLEX
beamline. This lack of transmission is not a fundamental
limitation and could be properly addressed in an optimized
setup. Likewise, it is exacerbated, in the present configuration,
by the large (2.5 mm) spacing between the beamlet
(reaching sub-ps bunching will eventually rely on shorter
spatial period on the cathode surface).

Nevertheless, the capabilities enabled by the
proposed technique offer a significant
improvement over setups previously considered.
Such a capability is of prime importance compared to setups previously investigated, e.g. in Ref.~\cite{SunPRL,HaPRL}, as it does not require any interceptive mask and is thereby compatible with, e.g., high-duty-cycle accelerators  contemplated as next generation accelerator-based light sources \cite{Zholents:IPAC2018-TUPMF010}. In addition, given that the shaping does not have any beam-intercepting hardware, the final distribution can, in principle, be dynamically controlled by varying the accelerator magnets, or altering the laser distribution at fast time scales [${\cal O}$(kHz)].

\begin{figure}[ttt!!!!]
\includegraphics[width=0.99\linewidth]{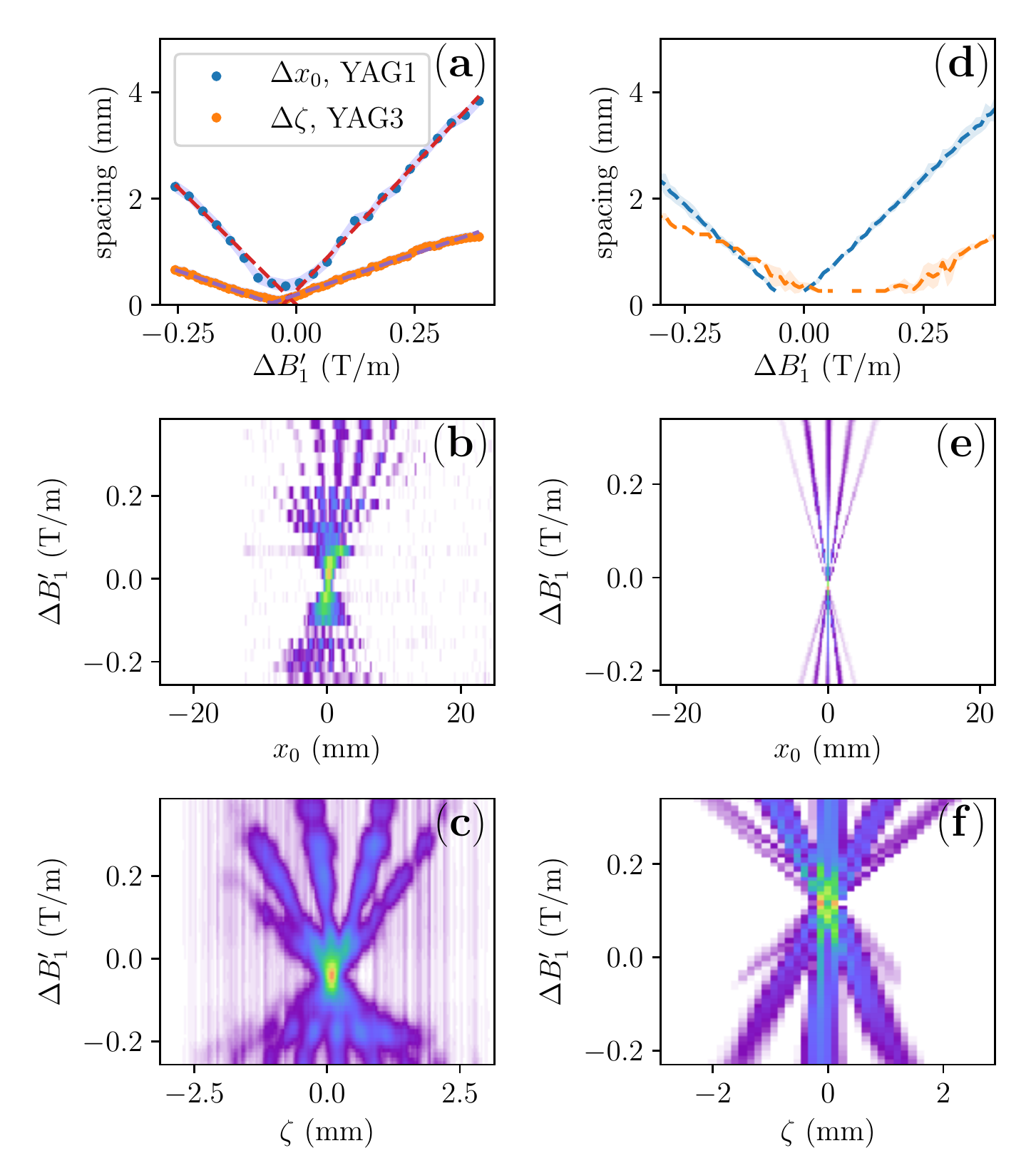}
 \caption{\label{fig:quadscan} Measured (left) and simulated (right) initial transverse $\Delta x_0$ and final longitudinal $\Delta \zeta$ beamlets separation (a,d) and corresponding horizontal (b,f) and longitudinal beamlet distributions (c,e) as a function of the change in magnetic-field gradient $\Delta B'_1$. 
 }
\end{figure}
\section{Dynamical control of the beam temporal modulation}
We first demonstrate the dynamical control of the final longitudinal bunch distribution using the quadrupole magnet Q1 to tune the final longitudinal spacing. To avoid some aberrations associated with our non-ideal setup, the MLA was rotated so that the symmetry axes of the electron-bunch array are aligned along the accelerator's horizontal and vertical directions [as displayed in Fig.~\ref{fig:trandist}(b)] and we selected only the central row of 5 beamlets with a mask [delineated by a box in Fig.~\ref{fig:trandist}(b)].  Figure~\ref{fig:quadscan}(a) shows the evolution of the initial horizontal ($\Delta x_0$) [resp. final longitudinal ($\Delta \zeta$)] pitch associated with the transversely [resp. longitudinally] segmented beam upstream [resp. downstream] of the TLEX. It is especially found that the transverse-to-longitudinal pitch ratio $\varrho\equiv \frac{\Delta \zeta}{\Delta x_0} = 0.34 \pm 0.03$ remains approximately constant as Q1 is varied. The {\sc impact-t} numerical simulations [also shown in Fig.~\ref{fig:quadscan}(d)] are in qualitative agreement with the experiment but the ratio $\varrho$ is found to be a function of $\Delta B'_1$ the change in field gradient compared to its nominal value $B'_1$. Such a discrepancy is attributed to an experimental uncertainty of the incoming beam-distribution parameters upstream of the quadrupole magnets. The evolution of the measured horizontal and longitudinal density distributions as a function of magnet Q1 setting respectively appears in Fig.~\ref{fig:quadscan}(b,c) while the corresponding simulations appear in Fig.~\ref{fig:quadscan}(e,f). 
The latter figures also demonstrate the beamlet temporal distribution downstream of the TLEX can be continuously tuned by using the quadrupole magnet to control the final LPS correlations. The beamlets temporally coalesce for $\Delta B'_1\simeq -0.05$~T/m corresponding to an upright longitudinal phase space (i.e. the beamlets are separated in energy only). For $\Delta B'_1\simeq 0.25$~T/m, the longitudinal spacing matches the initial transverse spacing at the cathode surface (of 2.5 mm) thereby realizing a one-to-one mapping one of the laser-distribution principal axis in the longitudinal (time) axis downstream of the TLEX beamline. Such an ability to continuously control the temporal distribution at the sub-ps scale is an appealing feature for, e.g., the generation of tunable THz radiation source, or the selective (resonant) excitation of electromagnetic wakefields in beam-driven acceleration techniques. \\

\begin{figure}[hhhhh!!!!]
\includegraphics[width=0.96\linewidth]{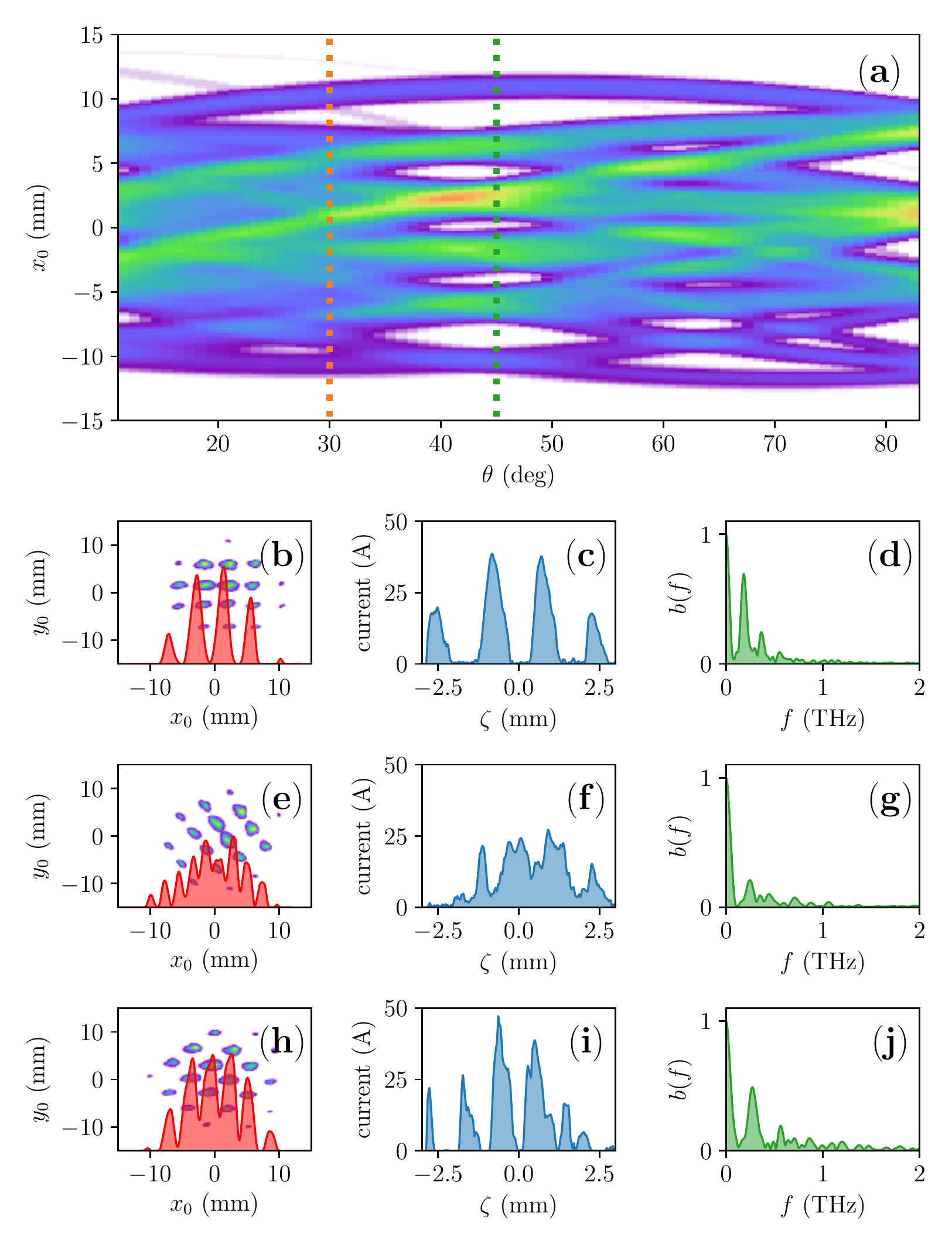}
 \caption{\label{rotation}  False color image of the horizontal beam distribution as a function of the MLA rotation angle at YAG1 (a). Measured peak-normalized beam transverse distribution (b,e,h) at YAG1 (false-color image) with superimposed horizontal projection (shaded-area traces), corresponding current distributions measured at YAG3 (c,f,i) with associated bunch form factors (d,g,j). The upper, middle, and lower rows respectively correspond to pattern-rotation angles on YAG 1 of $\theta=0$, $30$, and $45^{\circ}$.}
\end{figure}

An alternative method to control the beamlet separation consists in varying the laser-pattern angle at the photocathode surface. Such an approach also allows for the beamlets separation smaller than the array pitch~\cite{halavanauRotation} The MLA setup was configured for multi-beam generation and placed on a rotatable mount. A rotation of the MLA yields in rotation of the laser pattern on the photocathode surface. For the case of a square arrangement, the horizontal projection associated with the pattern distributions is altered as the rotation angle is scan within $\theta \in [0,90^{\circ}]$; see measured evolution of the horizontal projection densities in Fig.~\ref{rotation}(a). The latter figure also indicates that a smaller spatial period can be attained at selected rotation angles. 

We illustrate such a feature by comparing the distribution produced at the rotation angles $\theta=0$, 30, and $45^{\circ}$ with respect to horizontal axis with associated transverse distributions presented in Figure~\ref{rotation}(b,e,h). 
The resulting current distributions after transport through the  TLEX beamline appear in Figure~\ref{rotation}(c,f,i). For the nominal case of an upright pattern, the bunch separation is $\Delta \zeta \simeq 1.7$~mm while features with spatial scale below $500$~\textmu m are observed on the current profile for the case $\theta=30^{\circ}$. Likewise, Fig.~\ref{rotation}(d,g,j) gives the bunch form factor 
\begin{eqnarray}
b(f)\equiv \bigg|\int_{\infty}^{+\infty} \Lambda(\zeta) e^{-i2\pi f\zeta/c} d\zeta\bigg|,
\end{eqnarray}
 where $\Lambda$ is the normalized ($\int_{\infty}^{+\infty} \Lambda(\zeta) d\zeta =1$) charge distribution associated with each of the cases. The latter quantity confirms that non-zero angle provides higher-frequency spectral content. It should be noted that the observable features are limited by the resolution of our longitudinal diagnostics ($\Delta_\zeta \sim 200$~\textmu m) which therefore give an upper limit $f<1.5$~THz on the spectrum (this upper frequency would ultimately be controlled by the attainable single-beamlet duration). Additionally, the formation of smaller, e.g. sub-optical-scale, structures is ultimately limited by optical aberrations in the TLEX beamline~\cite{nanni}. It should however be pointed out that the spectral content achieved in the presented simple proof-of-principle experiment could be of interest, with minor improvements, to produce prebunched beams necessary for the generation of tunable coherent THz undulator radiation such as foreseen in pump-probe experiments planned at X-ray user facilities~\cite{THz}. The method could also serve as an alternative to a recently proposed sub-mm density-modulation scheme based on the interaction of the electron bunch with a laser~\cite{PhysRevAccelBeams.20.050701}.
Finally, it should be pointed out that the two methods explored for controlling the final current distribution (quadrupole magnet and rotation of the initial laser pattern) could be combined to provide a greater flexibility~\cite{halavanauRotation}.  \\

\section{Conclusion}
In Summary, we demonstrated space-to-time imaging of one of the projections associated with a photo-emission laser distribution impinging on a photocathode to the longitudinal (temporal) distribution of an electron beam. One promising application of the technique is the possible generation of narrowband electromagnetic radiation enhanced at the wavelength $\lambda_n \sim \Delta \zeta/n$ (where $n$ is an integer). Reaching ultrashort wavelength (high harmonic number $n$) will require the use of smaller laser pattern or nano-engineered cathodes capable of producing transversely-segmented beams~\cite{graves,li,anusorn}. In addition, attaining these finer-scale modulations will also require a thorough control of optical aberrations along the accelerator and in particular within the TLEX beamline~\cite{nanni, Shchegolkov:2014uxa}. The presented technique was shown to be versatile and allow for dynamical control of the final longitudinal bunch distribution. Although our proof-of-principle experiment focuses on the generation of an electron-bunch comb, any longitudinal distribution can in principle be realized owing to  recent advances in laser-shaping techniques. 

\section{Acknowledgments}
This work was supported by the US Department of Energy under contracts No. DE-SC0011831 and DE-SC0018656 with Northern Illinois University.  The AWA facility operation is funded through the U.S. Department of Energy contract No. DE-AC02-06CH11357. The work of A.H. and partially P.P. is supported by the US Department of Energy under contract No. DE-AC02-07CH11359 with the Fermi Research Alliance, LLC.

\end{document}